\documentclass[preprint,12pt]{elsarticle}
\usepackage{graphicx}
\usepackage{epstopdf}
\usepackage{subfig}
\usepackage{mathrsfs}
\usepackage{amssymb}
\usepackage[latin1]{inputenc}
\usepackage{float}
\usepackage{multirow}
\newtheorem{theorem}{Theorem}
\newtheorem{lemma}[theorem]{Lemma}
\newtheorem{corollary}[theorem]{Corollary}

\usepackage{hyperref}
\usepackage{xspace}
\usepackage{color}

\newcommand{\NP}{{\sf NP}\xspace}
\newcommand{\W}{{\sf W}\xspace}

\newcommand{\Nat}{{\mathbb N}}

\journal{Discrete Applied Mathematics}

\begin{document}
	
\begin{frontmatter}


\title{Maximum cuts in edge-colored graphs\tnoteref{t1}}

\tnotetext[t1]{An extended abstract of this paper was published in the \emph{Proc. of the IX Latin and American
Algorithms, Graphs and Optimization Symposium (\textbf{LAGOS}), \textbf{2017}}~\cite{sucupira2017maximum}.\\ This work was partially supported by the Brazilian agencies CNPq, CAPES and FAPERJ, and by French projects DEMOGRAPH (ANR-16-CE40-0028) and ESIGMA (ANR-17-CE40-0028).}

\author[uerj]{Luerbio Faria}
\ead{luerbio@cos.ufrj.br}

\author[ufrj]{Sulamita Klein}
\ead{sula@cos.ufrj.br}

\author[lirmm,UFC]{Ignasi Sau}
\ead{ignasi.sau@lirmm.fr}

\author[uff]{U\'everton S. Souza}
\ead{ueverton@ic.uff.br}

\author[uerj,ufrj]{Rubens~Sucupira}
\ead{rasucupira@gmail.com}

\address[uerj]{IME, Universidade Estadual do Rio de Janeiro, Rio de Janeiro, Brasil}
\address[ufrj]{IM, COPPE, Universidade Federal do Rio de janeiro, Rio de Janeiro, Brasil}
\address[lirmm]{CNRS, LIRMM, Universit\'{e} de Montpellier, Montpellier, France}
\address[UFC]{Departamento de Matem\'atica, Universidade Federal do Cear\'a, Fortaleza, Brasil}
\address[uff]{IC, Universidade Federal Fluminense, Niter\'oi, Brasil}

\begin{abstract}
The input of the \textsc{Maximum Colored Cut} problem consists of
a graph $G=(V,E)$ with an edge-coloring $c:E\to \{1,2,3,\ldots , p\}$ and  a positive integer $k$, and the question is whether $G$ has
a nontrivial edge cut  using at least $k$ colors.
The \textsc{Colorful~Cut} problem has the same input
but asks for a nontrivial edge cut using \emph{all} $p$ colors.
Unlike what happens for the classical \textsc{Maximum Cut} problem, we prove that both problems are
\NP-complete even on complete, planar, or bounded treewidth graphs. Furthermore,
we prove that \textsc{Colorful Cut} is \NP-complete even
when each color class induces a clique of size at most 3, but is trivially solvable when each color induces a $K_2$. On the positive side, we prove that \textsc{Maximum Colored Cut} is fixed-parameter tractable when parameterized by either $k$ or $p$, by constructing a cubic kernel in both cases.
\end{abstract}


\end{frontmatter}


\section{Introduction}\label{intro}

Given an edge-colored graph $G$ and an edge-set property $\Pi$, in maximum (minimum) colored/labeled $\Pi$ problems we are asked to find a subset of edges satisfying property $\Pi$ with respect to $G$ that uses the maximum (minimum) number of colors/labels. These problems have a lot of applications and have been widely studied in recent years:
\begin{itemize}

\item when $\Pi$ is the property of being a spanning tree of the input graph $G$, the {\sc Maximum (Minimum) Colored Spanning Tree} problems have been studied in \cite{broersma1997spanning,broersma2005paths,bruggemann2003local,chang1997minimum,hassin2007approximation,krumke1998minimum,voss2005applications};

\item when $\Pi$ is the property of being a path between two designated vertices of the input graph $G$, the {\sc Maximum (Minimum) Colored Path} problems have been studied in \cite{broersma2005paths,Carr:2000:RSC:338219.338271,hassin2007approximation,zhang2011approximation};

\item when $\Pi$ is the property of being a perfect matching of the input graph $G$, the {\sc Maximum (Minimum) Perfect Matching} problems have been studied in \cite{fellows2010parameterized,monnot2005labeled};

\item when $\Pi$ is the property of being a Hamiltonian cycle of the input graph $G$, the {\sc Minimum Colored Hamiltonian Cycle} problem has been studied in \cite{fellows2010parameterized};

\item when $\Pi$ is the property of being a edge dominating set of the input graph $G$, the {\sc Minimum Colored Edge Domination Set} problem has been studied in \cite{fellows2010parameterized};

\end{itemize}

In this work, we focus our studies on the complexity analysis of colored problems where $\Pi$ is the property of being an \emph{edge cut} of the input graph $G$.  More precisely, let $G=(V,E)$ be a simple graph with an edge coloring $c:E\to \{1,2,\ldots,p\}$, not necessarily proper. Given a proper subset $S\subset V$, we define the \emph{edge cut $\partial S$} as the subset of $E$ where the edges have one endpoint in $S$ and the other in $V\setminus S$. We represent by $c(\partial S)$ the set of colors that appear in $\partial S$, i.e., $c(\partial S)=\{c(e) \mid e\in \partial S\}$. The problem of finding a subset $S\subset V$ such that $|c(\partial S)|\leq |c(\partial T)|$ for every $T\subset V$ is called \textsc{Minimum Colored Cut}, and its decision version is stated as follows.

\bigskip

\vspace{.3cm}
\noindent	\fbox{
		\parbox{13cm}{
\noindent
{\sc \textsc{Minimum Colored Cut}} 

\noindent
\textbf{Instance}: A graph $G=(V,E)$ with an edge coloring $c:E \to \{1,2,\ldots,p\}$ and an integer $k>0$.

\noindent
\textbf{Question}: Is there a proper subset $S\subset V$ such that  $|c(\partial S)|\leq k$?
}
}

\bigskip

Associated with \textsc{Minimum Colored Cut} we have the \textsc{Minimum Colored $(s,t)$-Cut} problem, in which we are asked to find an edge cut that separates a given pair $s,t$ of vertices using as few colors as possible.

\smallskip

\vspace{.3cm}
\noindent	\fbox{
		\parbox{13cm}{
\noindent
{\sc \textsc{Minimum Colored $(s,t)$-Cut}} 

\noindent
\textbf{Instance}: A graph $G=(V,E)$ with an edge coloring $c:E \to \{1,2,\ldots,p\}$, a pair $s,t$ of vertices of $G$, and an integer $k>0$.

\noindent
\textbf{Question}: Is there a proper subset $S\subset V$ such that  $s\in S$, $t\notin S$ and $|c(\partial S)|\leq k$?
}
}

\vspace{.3cm}

Analogously, the problem of finding a subset $S\subset V$ such that $|c(\partial S)|\geq |c(\partial T)|$ for every $T\subset V$ is called \textsc{Maximum Colored Cut}, and its decision version is stated as follows.

\vspace{.3cm}
\noindent	\fbox{
		\parbox{13cm}{
\noindent
{\sc \textsc{Maximum Colored Cut}} 

\noindent
\textbf{Instance}: A graph $G=(V,E)$ with an edge coloring $c:E \to \{1,2,\ldots,p\}$ and an integer $k>0$.

\noindent
\textbf{Question}: Is there a proper subset $S\subset V$ such that  $|c(\partial S)|\geq k$?
}
}

\vspace{.3cm}

Note that the classical (simple) \textsc{Maximum Cut} problem~\cite{garey1976some} is the particular case of \textsc{Maximum Colored Cut} when  $c:E \to \Nat$ is an injective function. Therefore, for the \textsc{Maximum Colored Cut} problem we are interested in analyzing its complexity on graph classes $\mathcal{C}$ for which \textsc{Maximum Cut} is solvable in polynomial time.

\medskip

In addition, we are also interested in the complexity of determining if the input graph has a subset $S\subset V$ such that $|c(\partial S)|=p$, i.e., if there is an edge cut $\partial S$ using {\sl all} the colors; we call this problem \textsc{Colorful Cut}.

\vspace{.3cm}
\noindent	\fbox{
		\parbox{13cm}{
\noindent
{\sc \textsc{Colorful Cut}} 

\noindent
\textbf{Instance}: A graph $G=(V,E)$ with an edge coloring $c:E \to \{1,2,\ldots,p\}$.

\noindent
\textbf{Question}: Is there a proper subset $S\subset V$ such that  $|c(\partial S)|=p$?
}
}

\vspace{.3cm}


Complexity issues related to \textsc{Minimum Colored $(s,t)$-Cut} and \textsc{Minimum Colored Cut} have been widely investigated in recent years (cf.~\cite{Blin201466,bordini2017new,CoudertDPRV07,coudert2016combinatorial,fellows2010parameterized,Linqing,zhang2014,zhang2011approximation,zhang2016label,Zhang2018}).

Given an edge-colored graph $G$, the \emph{span} of a color of $G$ is the number of connected components of the subgraph induced by the edges of this color.

When the span of each color equals one, \textsc{Minimum Colored Cut} is equivalent to \textsc{Minimum Cut in Hypergraphs}, since each color class can be seen as a hyperedge. Therefore, it can be solved in polynomial time because, as observed in~\cite{Rizzi00}, the hypergraph cut function is symmetric and submodular~(cf.~\cite{queyranne1998minimizing,Rizzi00}). Coudert et al.~\cite{CoudertDPRV07} showed that \textsc{Minimum Colored Cut} can be solved in polynomial time when the number of edges of each color is bounded by a given constant, but  \textsc{Minimum Colored $(s,t)$-Cut} is \NP-hard even when each color contains at most two edges. Blin et al.~\cite{Blin201466} presented a randomized algorithm that returns an optimal colored cut of $G$ with probability at least $(|V|^{2k})^{-1}$, where $k$ is the maximum span of $G$. Sevaral approximation and hardness results for  \textsc{Minimum Colored $(s,t)$-Cut} are presented in~\cite{Linqing,zhang2014,zhang2011approximation,Zhang2018}. Zhang and Fu~\cite{zhang2016label} showed that \textsc{Minimum Colored $(s,t)$-Cut} is \NP-hard even if the maximum length of any path is equal to two; and when restricted to disjoint-path graphs, \textsc{Minimum Colored $(s,t)$-Cut} can be solved in polynomial time if the number of edges of each color is at most two. Regarding the parameterized complexity of \textsc{Minimum Colored $(s,t)$-Cut}, Fellows et al.~\cite{fellows2010parameterized} showed that  the problem is $\W[2]$-hard when parameterized by the number of colors of the solution, and  $\W[1]$-hard when parameterized by the number of edges of the solution.  Coudert et al.~\cite{coudert2016combinatorial} showed that \textsc{Minimum Colored Cut} can be solved in time $2^k\cdot n^{O(1)}$, where $k$ is the number of colors with span larger than one.

The main goal of this work is to present a complexity analysis of \textsc{Maximum Colored Cut}, which, to the best of our knowledge, was missing in the literature. As \textsc{Colorful Cut} is a particular case of \textsc{Maximum Colored Cut}, our hardness results deal with \textsc{Colorful Cut}, while the tractable cases will be presented for \textsc{Maximum Colored Cut}.

\smallskip

The remainder of the article is organized as follows. In Section~\ref{sec:NPc} we provide several \NP-completeness results for restricted versions of \textsc{Colorful Cut}, and in Section~\ref{sec:FPT} we present cubic kernels for \textsc{Maximum Colored Cut} parameterized either by $p$ or by $k$. We use standard graph-theoretic notation; see~\cite{Die05} for any undefined notation. For the basic definitions of parameterized complexity, such as fixed-parameter tractability, $\W[2]$-hardness, para-\NP-hardness, or (polynomial) kernelization, we refer the reader to~\cite{CyganFKLMPPS15,DF13}.

\section{\NP-completeness results for \textsc{Colorful Cut}}
\label{sec:NPc}

Hadlock~\cite{hadlock1975finding} proved that (simple) \textsc{Maximum Cut} is polynomial-time solvable on planar graphs. In this section we prove, among other results, the \NP-completeness of \textsc{Colorful Cut} on a particular subclass of planar graphs. We start with general planar graphs, and then we discuss how the construction can be modified to get stronger hardness results.

\begin{theorem}\label{planares}
\textsc{Colorful Cut} is \NP-complete on planar graphs.
\end{theorem}

\noindent {\bf Proof. }
Let $I=(U,C)$ be an instance of \textsc{3-sat}. We construct in polynomial time a planar instance $G=(V,E)$ with an edge coloring $c$ such that $I=(U,C)$ is satisfiable if and only if $(G,c)$ has an edge cut using all colors of $c$.

\begin{figure}[h!]
\centering
	\subfloat
			[]
			{
\includegraphics[width=.46\textwidth]{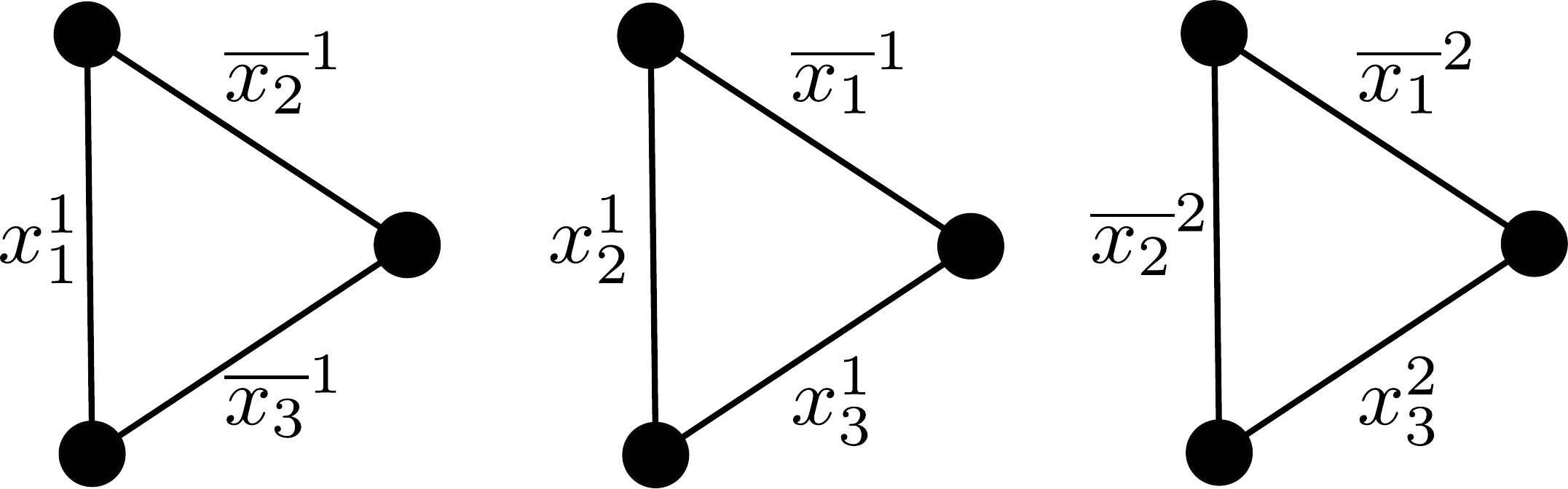}
			}
\hspace{0.5cm}
	\subfloat
			[]
			{
\includegraphics[width=.5\textwidth]{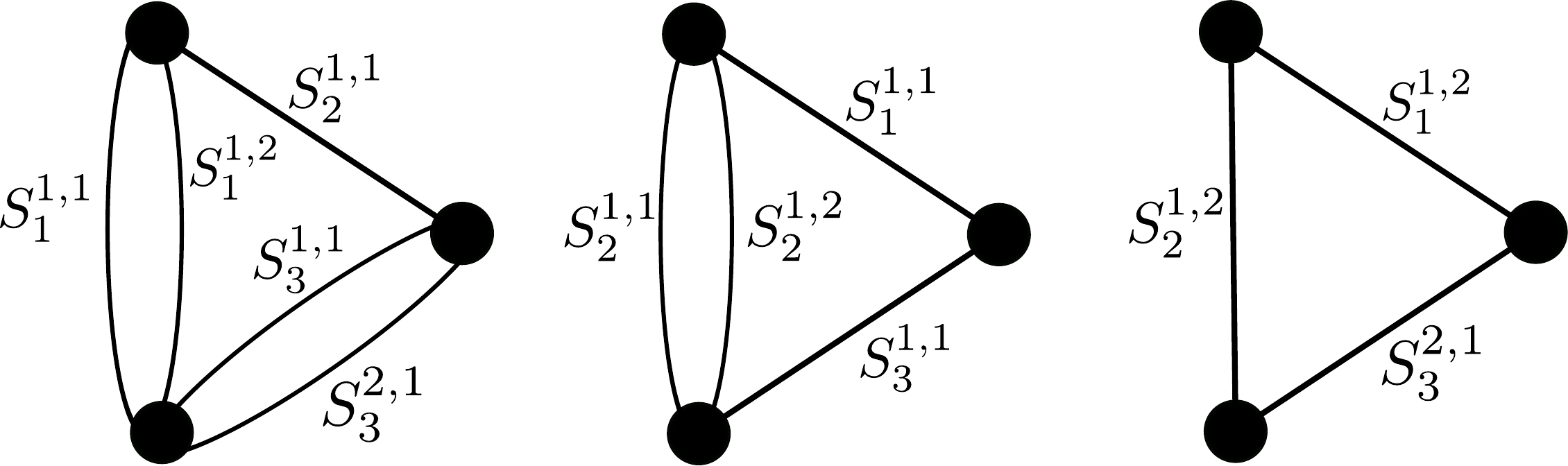}
			}
			
\caption {\footnotesize Graphs $G'$ and $G^c$ associated with $(x_1\vee \overline{x_2}\vee \overline{x_3})$ $\wedge (\overline{x_1}\vee x_2 \vee x_3)\wedge (\overline{x_1}\vee \overline{x_2}\vee x_3)$.}\label{gadget3sat}
\end{figure}

With each clause $c_j=(x \vee y \vee z)\in C$ we associate a $K_3$ where each edge is labeled by a literal of $c_j$ together with its occurrence in $C$,
obtaining a graph $G'$ with $m$ connected components (see Figure~\ref{gadget3sat}(a)), each of them associated with a clause of $C$. Starting with $G'$, we construct a multigraph $G$ equipped with a coloring $c$, for short $G^c$, where each literal in a clause of $C$ is associated with a colored (multi)edge as follows:

\begin{itemize}
	\item For each pair $\{x_i^j,\overline{x_i}^k\}$, where the integers $j$ and $k$ represent occurrences of the literals $x_i$ and $\overline{x_i}$ in the clauses of $C$, respectively, create a color denoted by $S_i^{j,k}$.
	\item The edge labeled with $x_i^j$ in $G'$ is replaced with parallel edges colored with $S_i^{j,k}$ in $G^c$ for all $k$. Analogously, the edge labeled by $\overline{x_i}^k$ in $G'$ is replaced with parallel edges in $G^c$ colored with $S_i^{j,k}$ for all $j$.
\end{itemize}

Figure~\ref{gadget3sat} illustrates the graphs $G'$ and $G^c$ associated with an instance $I$.

Without loss of generality, we may assume that all variables have both po\-si\-ti\-ve and ne\-ga\-ti\-ve literals in $I$ (if not, the clauses containing such variables are trivially satisfiable and can be removed). From a truth assignment $A_I$ of $I$, we can construct a colorful cut of $G^c$ as follows. For each clause $c_j$ of $C$,  pick arbitrarily one edge $\{v,w\}$ corresponding to a true literal of $C$. Then, put $v$ and $w$ in the same part of the partition, leaving the remaining vertex of the clause in the other part. This procedure gives a cut using all colors. Indeed, for each true literal $x_i^j$, there is at least one false literal $\overline{x_i}^k$ that places the color $S_i^{j,k}$ in the cut.

\begin{figure}[h!]
	\centering
	\includegraphics[width=.67\textwidth]{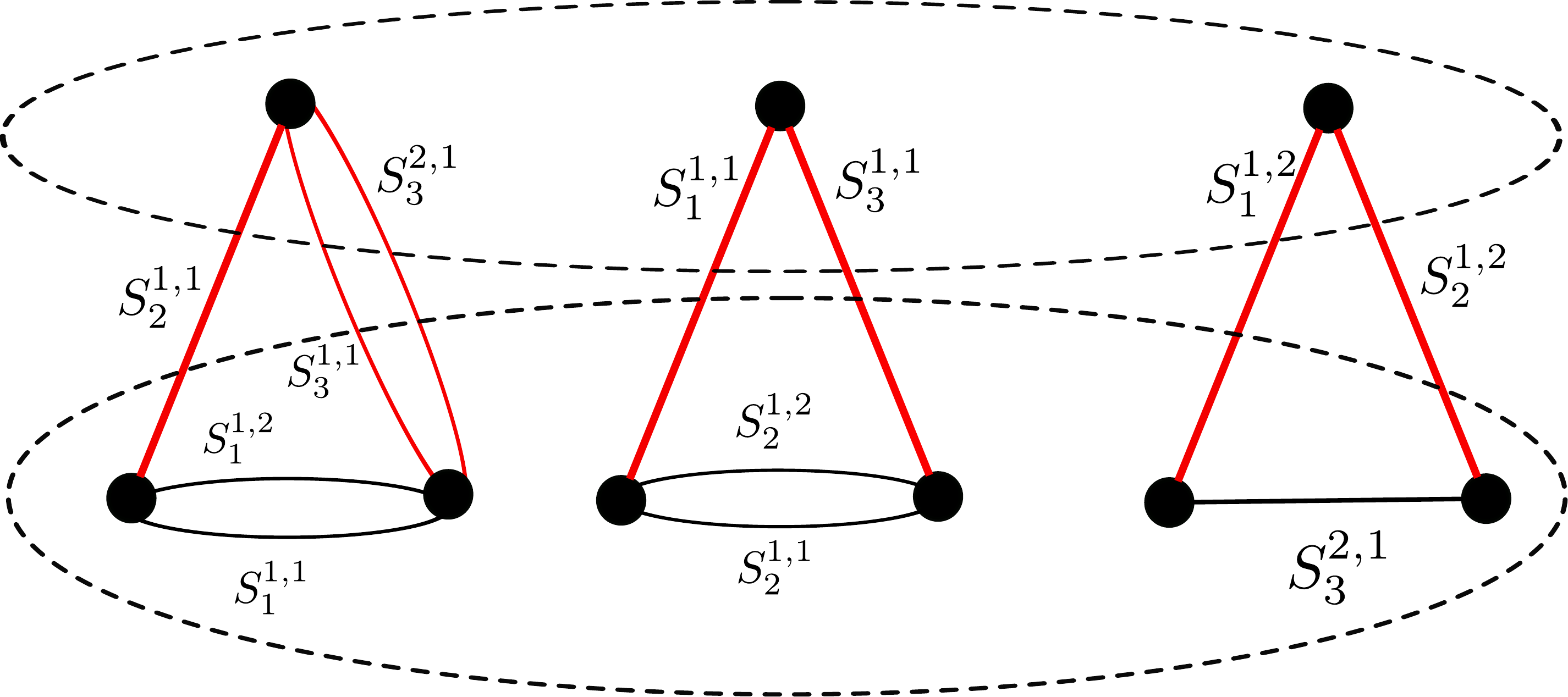}
	\caption{\footnotesize Edge cut associated with the instance $(x_1\vee \overline{x_2}\vee \overline{x_3})\wedge (\overline{x_1}\vee x_2 \vee x_3)\wedge (\overline{x_1}\vee \overline{x_2}\vee x_3)$ with truth assignment $x_1=x_2=x_3=1$.}\label{multicolpart}
\end{figure}

Figure~\ref{multicolpart} illustrates a colorful cut for the multigraph $G^c$ associated with the instance of Figure~\ref{gadget3sat}, considering the truth assignment $A_I=x_1=x_2=x_3=1$ as previously described.

Conversely, suppose that $G^c$ has a colorful cut. Without loss of generality, we may assume that each $K_3$ has a cut edge Indeed, as the cut has all the colors of the edge coloring, if there is some $K_3$ in a part of the partition, we can choose any vertex of this clique and place this vertex in the other part, without prejudice, because all the colors are still in the cut.
Beginning with this cut, we construct a truth assignment $A_I$ that satisfies $I$, putting $x_i=1$ if at least one of the edges associated with some $x_i^j$ is inside a part of the partition, and $x_i=0$ otherwise. Note that this assignment is well-defined: there is no pair of literals $\{x_i^j,\overline{x_i}^k\}$ such that the edges corresponding to both literals are inside a part of the partition, otherwise the color $S_i^{j,k}$ is missing and the cut does not contain all colors. Besides that, each $K_3$ has the edges corresponding to some literal inside a part of the partition, which defines a truth assignment for $I$.

Finally, we can transform $G^c$ into a simple graph replacing each edge $\{v,w\}$ colored with $S_i^{j,k}$ by a path $\{v,x,y,w\}$ such that $c(\{x,y\})=S_i^{j,k}$ and the remaining edges of this path receive new different colors.

\begin{figure}[h!]
	\centering
	\includegraphics[width=.63\textwidth]{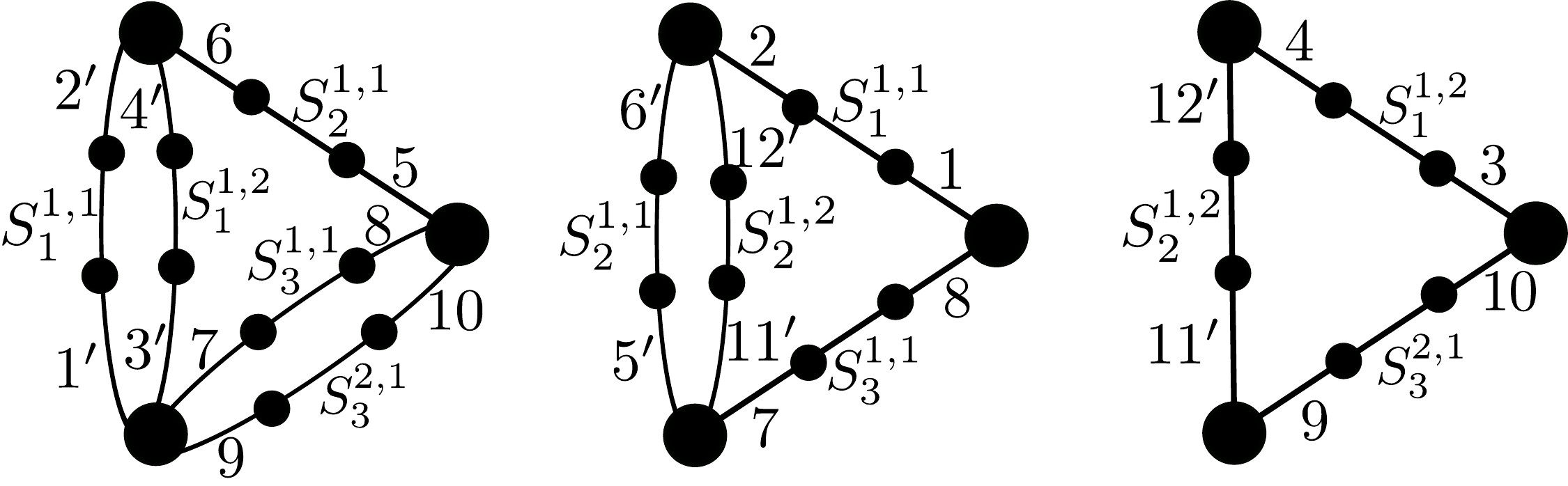}
	\caption{\footnotesize Graph $H$ associated with $G^c$.}\label{fig:gadgetsimples}
	
\end{figure}

Figure~\ref{fig:gadgetsimples} illustrates the simple graph $H$ obtained from $G^c$, where the new colors are labeled with numbers. It is not difficult to see that the multigraph $G^c$ has a colorful cut if and only if the associated graph $H$ has a colorful cut.\qed

\begin{figure}[ht]
	\centering
	\includegraphics[width=.92\textwidth]{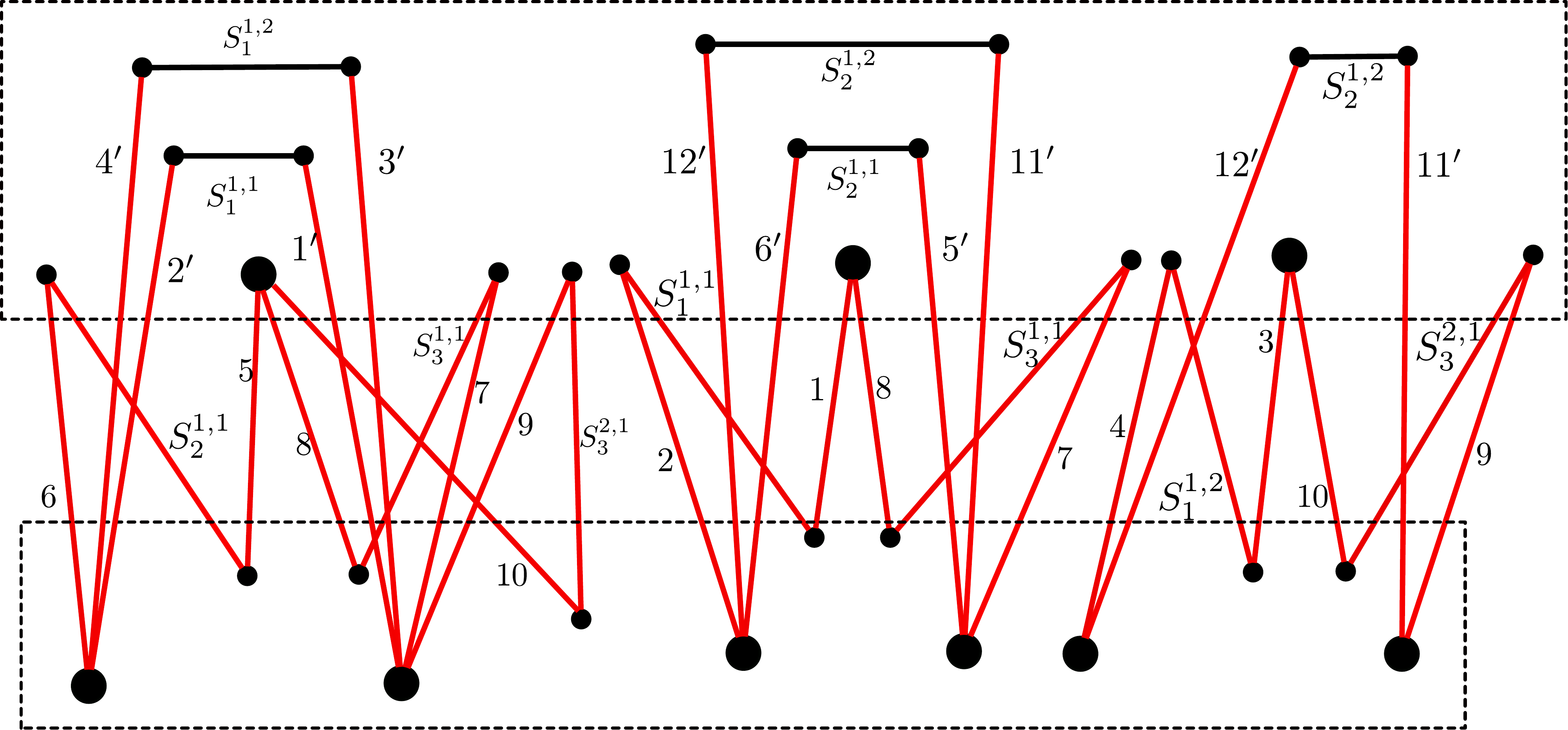}
	\caption{\footnotesize Colorful cut of $H$.}\label{fig:gadgetsimplescort}
	
\end{figure}

Figure~\ref{fig:gadgetsimplescort} illustrates a colorful cut of $H$.

Several \NP-hard problems, such as (simple) \textsc{Maximum Cut}, are polynomial-time solvable on bounded treewidth graphs~\cite{Bodlaender1994}. An important class of graphs that belong to the intersection of planar graphs and graphs of bounded treewidth is the class of  \emph{$K_4$-minor-free} graphs~\cite{bodlaender1998partial, duffin1965topology}.

A graph $G$ is called \emph{series-parallel} if it can be obtained from a $K_2$ by applying a sequence of operations, where each operation is either to duplicate an edge (i.e., replace an edge with two parallel edges) or to subdivide an edge (i.e., replace an edge with a path of length two). A graph $G$ is $K_4$-minor free if and only if each 2-connected component of $G$ is a series-parallel graph~\cite{bodlaender1998partial,duffin1965topology}.

$K_4$-minor-free graphs are planar because $K_5$ and $K_{3,3}$ have $K_4$ as a minor. In addition, $K_4$-minor-free graphs are the graphs with treewidth two~\cite{bodlaender1998partial}.

We can modify the graph obtained in the construction presented in the proof of Theorem~\ref{planares} in order to obtain the following corollary.

\begin{corollary}\label{cor:hard-planar}
\textsc{Colorful Cut} remains \NP-complete even when the input graph $G$ satisfies simultaneously  the following properties:
\begin{enumerate}
\item $G$ is $K_4$-minor-free.
\item $G$ is connected.
\item $G$ has maximum degree three.
\item Each color class of $G$ contains at most two edges.
\end{enumerate}
\end{corollary}

\noindent {\bf Proof. }
Let $H$ be an instance of \textsc{Colorful Cut} constructed as described in the  proof of Theorem~\ref{planares}. First observe that each connected component of $H$ (clause gadget) can be obtained from a $K_2$ by either duplicating an edge or subdividing an edge. Therefore, each connected component of $H$ is series-parallel.

In order to make the graph $H$ be connected and with bounded degree just create a binary tree $T$ with $m$ leaves and add edges by connecting a vertex with maximum degree of each gadget clause of $H$ to a distinct leaf of $T$. Assigning a new distinct color for each edge previously created, it holds that $H$ is $K_4$-minor-free and each color class of $H$ contains at most two edges (as in the proof of Theorem~\ref{planares}). Finally, each vertex $v$ of degree 4 can be replaced by a $P_3$ where each pedant vertex is neighbor of two vertices that were adjacent to $v$ and came from the same edge in $G'$ (the edges of these $P_3$'s also get new colors).

As the set of edges that we add in the graph induces a tree having a new color for each edge, it is easy to see that the modified graph has a colorful cut if and only if the original graph has a colorful cut. \qed

\bigskip

Clearly, every bipartite graph has a colorful cut. Thus, it is natural to ask about the complexity of the problem on graphs with small odd cycle transversal.

\begin{corollary}
\textsc{Colorful Cut} remains \NP-complete even when the input graph $G$ has odd cycle transversal number one.
\end{corollary}

\noindent {\bf Proof. }
It is enough to pick a vertex of each gadget of $(G,c)$, constructed as described in the  proof of Theorem~\ref{planares}, and identify them into a single vertex.
\qed

\medskip

Note that \textsc{Maximum Cut} is trivial on complete graphs, and that it is polynomial time solvable on cographs~\cite{Bodlaender1994}. By adding a new vertex and edges colored with a new color, we can construct a hard instance in order to show the \NP-completeness of \textsc{Colorful Cut} on complete graphs.

\begin{theorem}\label{maxcompletos}
\textsc{Colorful Cut} is \NP-complete on complete graphs.
\end{theorem}

\noindent {\bf Proof. }
Given an instance $(G,c)$ of \textsc{Colorful Cut}, we create another instance
$(G',c')$ such that $G'$ is a clique as follows. Start from $(G,c)$, add all the missing edges to $G$, add a new vertex $v$ adjacent to all the vertices of $G$, add give to the edges in $E(G') \setminus E(G)$ the same color, different from the colors appearing in $E(G)$. Clearly, this new color appears in all the maximum colored cuts of $G'$, and therefore $(G',c')$ has a colorful cut if and only if $(G,c)$ has one.
\qed

\medskip

Note that if each color class of a graph $G$ induces a $K_2$, then $G$ has a colorful cut if and only if $G$ is bipartite, which can be decided in polynomial time. The next result shows that this is best possible, in the sense that \textsc{Colorful Cut}  is  \NP-complete when each color class induces either a $K_2$ or a $K_3$.

\begin{theorem} \label{teoluerbio}
\textsc{Colorful Cut} is \NP-complete when each color class induces a clique of size at most three.
\end{theorem}

\noindent {\bf Proof. }
For this proof we use a reduction from \textsc{Not All Equal 3-sat (nae 3-sat)}, which is \NP-complete~\cite{thomasschaefer}. Let $I=(U,C)$ be an instance of \textsc{nae 3-sat} such that $U=\{u_1,u_2,\ldots,u_n\}$ and $C=\{c_1,c_2,\ldots,c_m\}$.

The construction of an instance $(G,c)$ is given by the following procedure:

\begin{itemize}
	\item For each clause $c_j=(x,y,z)\in C$, construct a clique $\{(x)_j,(y)_j,(z)_j\}$ with all the edges colored with color $j$.
	
	\item For each variable $u_i\in U$, add two new vertices $a_i$ and $b_i$ to $V$, such that $a_i$ is only adjacent to all positive occurrences of $u_i$, and $b_i$ is only adjacent to all negative occurrences of the same variable.

\item For each variable $u_i\in U$, add an edge joining the vertices (in the clause cliques) corresponding to the first positive occurrence and the first negative occurrence of $u_i$.

	\item Excluding the edges of the clause cliques, all other edges are colored with new different integers strictly greater than $m$.	
\end{itemize}

\begin{figure}[h!]
\centering
\includegraphics[width=.65\textwidth]{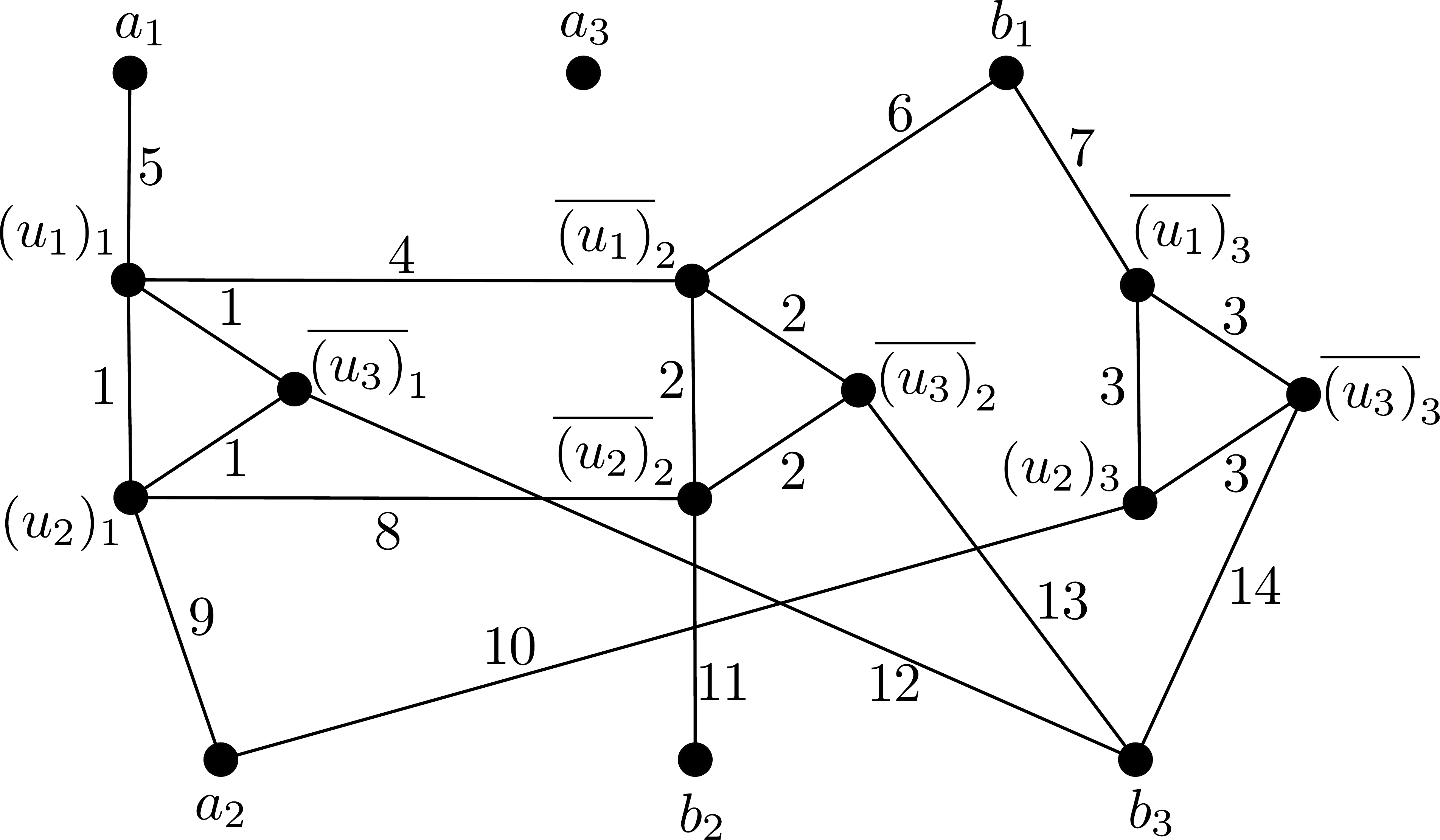}
\caption{\footnotesize Graph associated with the formula $(u_1\vee u_2\vee \overline{u_3}) \wedge(\overline{u_1}\vee \overline{u_2}\vee\overline{u_3}) \wedge(\overline{u_1}\vee u_2 \vee \overline{u_3})$.} \label{fig:gadgetluerbio}
\end{figure}

Figure~\ref{fig:gadgetluerbio} illustrates the instance $(G,c)$ of \textsc{Colorful Cut}  associated with an instance $I=(U,C)$ of \textsc{Not All Equal 3-sat}.

At this point, it is not difficult to see that $I=(U,C)$ is a satisfiable instance of \textsc{nae 3-sat} if and only if $(G,c)$ has a colorful cut. Indeed, suppose first that $I=(U,C)$ is a satisfiable instance of \textsc{nae 3-sat}, and let $\eta$ be a truth assignment of $U$ that satisfies $I$. A colorful cut $\partial S$ in $G^c$ is obtained as follows: if $u_i=1$, then put  all its positive occurrences together with $b_i$ in $S$, and put all its negative occurrences together with $a_i$ in $V\setminus S$. By construction, all the colors greater than $m$ are in the cut $\partial S$. Furthermore, each of the colors $j\leq m$ is in the cut because in each clause there is always a true and a false occurrence.

Conversely, suppose that $(G,c)$  contains a colorful cut $\partial S$. All the  clause cliques have vertices in different parts of the partition, because the colors $j$ with $1\leq j \leq m$ only appear in those clique edges. Thus we can produce a truth assignment $\eta$ by setting to true to those literals corresponding to the clique vertices $\{x_j,y_j,z_j\}$ that belong to $S$, and by setting to false otherwise. This is a consistent truth assignment because the edge joining the first positive and negative occurrences of the variable $u_i$ (if any) must be a cut edge, that is, its exclusive color must be in the cut, which means that those occurrences must be in different parts of the partition, thus having opposite truth assignments. As all positive occurrences of $u_i$ are adjacent to the vertex $a_i$ and those edges have pairwise different colors presented in the cut, it forces all positive occurrences of $u_i$ to be in the same part of the partition, receiving the same truth assignment. Analogously, we can prove that all negative occurrences of $u_i$ must be in the same part of the partition, getting the same truth assignment.

\begin{figure}[h!]
	\centering
	\includegraphics[width=.8\textwidth]{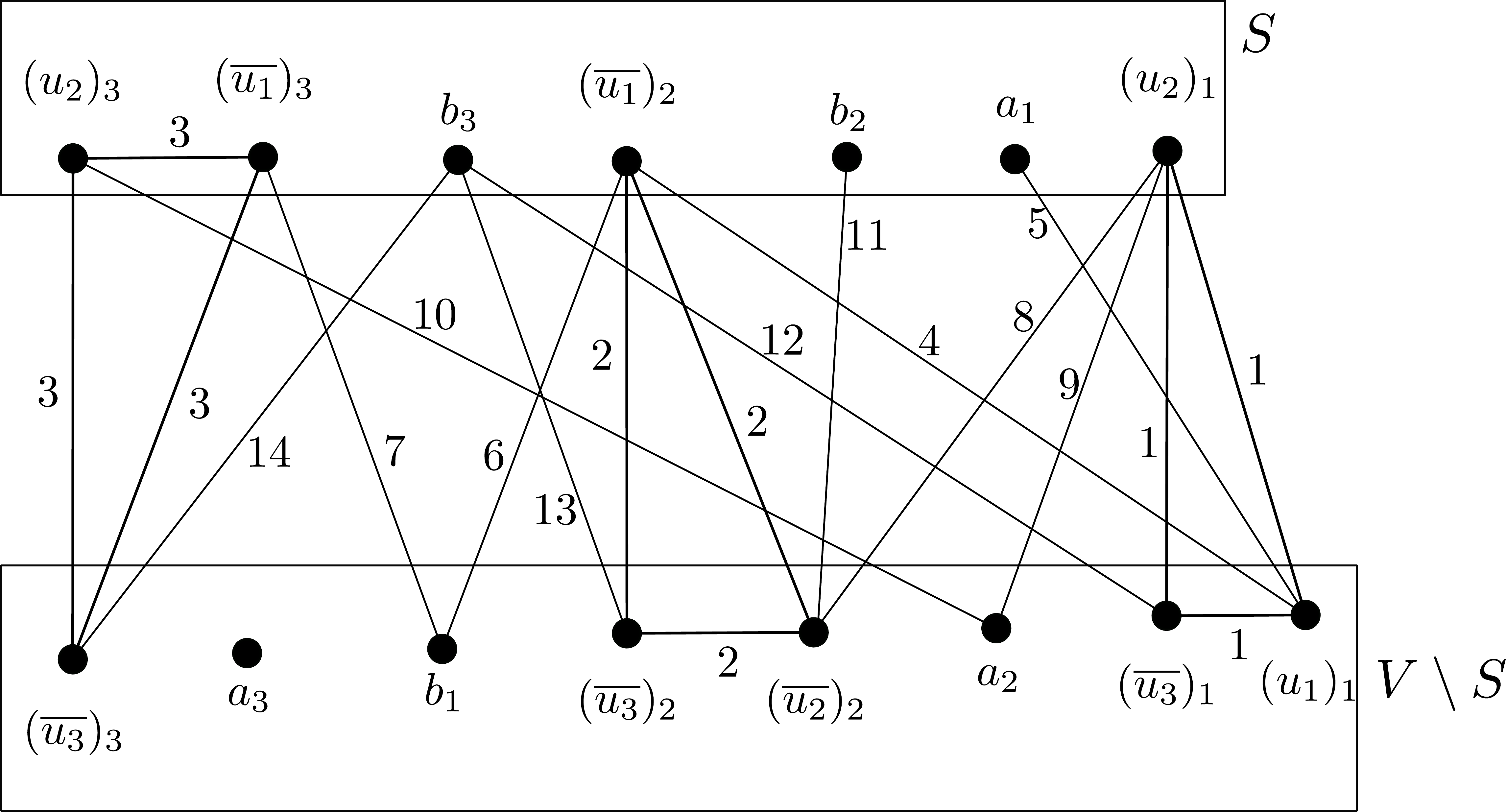}
	\caption{\footnotesize Colorful cut of $G^c=(V,E,f)$ from Figure~\ref{fig:gadgetluerbio} corresponding to the truth assignment $\overline{u_1}=u_2=u_3=1$.}\label{gadlupart}
\end{figure}

Figure~\ref{gadlupart} illustrates the colorful cut of $(G,c)$ from the \textsc{nae 3-sat} instance $I=(U,C)$ with $U=\{u_1,u_2,u_3\}$ and $C=\{(u_1\vee u_2\vee \overline{u_3}),(\overline{u_1}\vee \overline{u_2}\vee\overline{u_3}),(\overline{u_1}\vee u_2 \vee \overline{u_3})\}$, satisfying the truth assignment $\overline{u_1}=u_2=u_3=1$.
\qed

\medskip

\section{Polynomial kernelization of \textsc{Maximum Colored Cut}}
\label{sec:FPT}

From the results presented in Section~\ref{sec:NPc} it follows that \textsc{Maximum Colored Cut} is {\em para-\NP-hard} (see~\cite{CyganFKLMPPS15,DF13}) parameterized by any of these parameters: treewidth, neighborhood diversity, genus, de\-ge\-ne\-ra\-cy, odd cycle transversal number, $p-k$, and several combinations of such parameters. In contrast to these results, next we show the fixed-parameter tractability of  \textsc{Maximum Colored Cut} when parameterized by either $k$ or $p$, by means of the existence of a polynomial kernel.

\begin{theorem}\label{lem:max-color-kernel}
 \textsc{Maximum Colored Cut} admits a cubic kernel parameterized by the number of colors. 
\end{theorem}

\noindent {\bf Proof. }
First recall that a cut of a graph $G$ is a bipartite subgraph of $G$. The following claim is an easy fact.

\medskip

\noindent \textbf{Claim 1.} Let $H=(V_1,V_2,E)$ be a bipartite graph having $\beta$ edges and no isolated vertices. The maximum number of edges having endpoints in the same part that can be added to $H$ is $2{\beta \choose 2}$, corresponding to the case where $E$ induces a matching.

\medskip

Now, suppose that  $\lambda$ is the maximum number of colors in a cut of $G=(V,E)$ and let $S\subset V$ be a set such that $|c(\partial S)|=\lambda$.  Forming a bipartite graph $H$ by selecting exactly one edge of each color class in $[S,V\setminus S]$, by Claim~1 it follows that any color class that is not in $H$ has at most $2{\lambda \choose 2}$ edges, otherwise $\lambda$ would not be maximum.
Let $E_i\subseteq E$ be the set of edges colored with color $i$. As $\lambda\leq p$, if $|E_i| > 2{p \choose 2}$, then color $i$ appears  in any maximum colored cut. Such a property gives us the following reduction rule:

\begin{itemize}
\item[$\star$] If for some color $i$,  $|E_i| > 2{p \choose 2}$, decrease by one the number of colors and replace $G$ by $G[E\setminus E_i]$.
\end{itemize}

The exhaustive application of Rule~$\star$ yields a kernel of size $O(p^3)$.
\qed

\medskip

Before our last result, we need the following lemma.

\begin{lemma}\label{p/2}
Any simple graph $G=(V,E)$ with an edge coloring $c:E \to \{1,2,\ldots,p\}$ has an edge cut $\partial S$ such that $|c(\partial S)|\geq \frac{p}{2}$.
\end{lemma}

\noindent {\bf Proof. }
Let $G'$ be an uncolored graph obtained from $G$ by keeping one arbitrary edge from each color. Then the lemma follows by applying to $G'$  the well-known property that any graph with at least $m$ edges contains a bipartite subgraph with at least $\frac{m}{2}$ edges~\cite{erdos}.
\qed

\medskip

\begin{corollary}\label{lem:max-color-kernel-colors2}
 \textsc{Maximum Colored Cut} admits a cubic kernel parameterized by the cost of the solution.
\end{corollary}

\noindent {\bf Proof. }
If $k \geq p/2$, by Lemma~\ref{p/2} we conclude that we are dealing with a \textsc{Yes}-instance. Otherwise, $k < p/2$, and applying Rule~$\star$ exhaustively yields a kernel of size $O(k^3)$.
\qed

\bibliographystyle{abbrv}
\bibliography{qualificacao}

\end{document}